# Silicene Like Domains on IrSi$_3$ Crystallites

Dylan Nicholls, Fatima, Deniz Çakır, Nuri Oncel*

Department of Physics and Astrophysics

University of North Dakota

Grand Forks, North Dakota, 58202, USA

*Corresponding Author

Email: nuri.oncel@und.edu

**Abstract:** Recently, silicene, the graphene equivalent of silicon, has attracted a lot of attention due to its compatibility with Si-based electronics. So far, silicene has been epitaxy grown on various crystalline surfaces such as Ag(110), Ag(111), Ir(111), ZrB$_2$(0001) and Au(110) substrates. Here, we present a new method to grow silicene via high temperature surface reconstruction of hexagonal IrSi$_3$ nanocrystals. The h-IrSi$_3$ nanocrystals are formed by annealing thin Ir layers on Si(111) surface. A detailed analysis of the STM images shows the formation of silicene like domains on the surface of some of the IrSi$_3$ crystallites. We studied both morphology and electronic properties of these domains by using both scanning tunneling microscopy/spectroscopy and first-principles calculation methods.

**Introduction:**

Graphene has attracted a lot of attention for its unique properties, which include excellent electrical/thermal conductivity and strong mechanical strength. [1,2] The immediate alternatives for graphene are from the other group IV elements in the periodic table with similar electron configuration, i.e. Silicon (Si) and Germanium (Ge). [3] Although all three elements have four electrons in their outermost s- and p-orbitals, energetically the most favorable crystal structure of Si and Ge is the diamond structure.[4] Therefore, for Si and Ge, graphite-like allotropes (hereafter referred as silicene and germanene) don't exist in nature. Hence silicene and germanene have to be synthesized. The term "silicene" was introduced by G. G.

Guzman-Verri and L. C. Lew Yan Voon in 2007 to refer to two-dimensional structure of silicon atoms in honeycomb arrangement.[5] Takeda and Shiraishi were the first to predict the possibility of a stable single sheet of Si.[6] In that paper, the authors predicted that the silicene will not be flat but puckered. Further theoretical calculations on free-standing silicene showed that this two-dimensional system can be stable with properties similar to graphene such as linear dispersion with Dirac cone at the corner of Brillouin zone.[7, 8, 9] Theoretical studies also showed that the spin-orbit coupling strength is much larger in silicene than graphene which can make quantum spin Hall Effect detectable experimentally.[10]

Silicene's novel properties predicted by theorists and its relatively easy integration into the current semiconductor technology motivated some experimental groups to start working on growing silicene. The very first silicene was grown on Ag(110) and Ag(111) surfaces.[11-20] Recently, it was shown that silicene can be grown on Ir (111)[21], $ZrB_2$ (0001)[22], Au (110)[23] substrates. One of the most studied silicene system is silicene/Ag(111). Due to the interaction between silicene and substrate, the bucking pattern of Si atoms is rearranged resulting in various superstructures such as $(4 \times 4)$, $(\sqrt{13} \times \sqrt{13})R13.9°$, $(2\sqrt{3} \times 2\sqrt{3})$, [with respect to Ag(111) surface lattice] and $(3 \times 3)$ [with respect to silicene (1×1)].[24-26] All these superstructures show slightly different honeycomb configurations. ARPES measurements on $4 \times 4$-silicene along Ag $\overline{\Gamma} - \overline{K}$ direction through the silicene $\overline{K}$ point shows a downward dispersing branch of the honeycomb silicene bands. The dispersion is similar to the dispersion of graphene around the $\overline{K}$ point indicating that electrons behave as massless Dirac Fermions.[27] ARPES measurements on $\sqrt{3} \times \sqrt{3}$-silicene exhibit $\wedge-$ and $\vee-$ shaped linear π and π* silicene bands at the $\overline{\Gamma}_0$ point along $\overline{\Gamma} - \overline{K}_{silicene}$ direction.[28]

Another method to grow silicene or germanene is to use high temperature surface reconstruction of metal silicides/germanides.[29, 30] In this paper, we show that the deposition of a few monolayers of Ir on Si(111) followed by annealing at 750 °C leads to the formation of $IrSi_3$ nanocrystals. Scanning tunneling microscopy (STM) images of these crystallites show a surface reconstruction with a buckled honeycomb structure that resembles previously reported silicene like regions on $MoSi_2$ crystallites supported on Si(001).

**Experiment**

The Si(111) samples used in the STM experiments were cut from nominally flat 76.2 mm by 0.38 mm, single side-polished n-type (phosphorous doped) wafers. The samples were mounted on molybdenum holders and contact of the samples to any other metal during preparation and experiment was carefully avoided. The STM/STS studies have been performed by using an ultra-high vacuum system with a base pressure of $2 \times 10^{-10}$ mbar equipped with an Omicron Variable Temperature STM and a LK technologies RVL2000 LEED-Auger system. Before introducing Si(111) samples into the UHV chamber, samples were washed with isopropanol and dried under the flow of nitrogen gas. Si(111) samples were degassed extensively and then flash-annealed at 1250 °C. The sample temperature was measured with a pyrometer. The quality of the clean Si(111) samples was confirmed both with LEED and STM prior to Ir deposition. Ir was deposited on the clean Si(111)- $7 \times 7$ surface kept at room temperature from a current heated Ir wire (99.9 %) with a standard deposition rate of $2.8 \times 10^{-4}$ nm/s. Then the sample was annealed at 750 °C to form IrSi$_3$ nanocrystals. All the STM experiments were performed at room temperature. I-V curves measured at all points of the image while measuring high resolution STM images of the surface. The measured I-V curves were then averaged. The local density of states curves (LDOS-(dI/dV)/(I/V)) were calculated out of the I-V curves.

**Computational Details**

We performed first-principles calculations based on density functional theory as implemented in the Vienna Ab-initio Simulation package (VASP).[31,32] We described the ions by employing PAW pseudopotentials for atomic core region.[33] The generalized gradient approximation (GGA) was used for the exchange-correlation functional within the Perdew-Burke-Ernzerhof (PBE) formulation.[34] The kinetic energy cut-off for plane wave expansion was set to 400 eV. We used $3 \times 3 \times 1$ and $5 \times 5 \times 1$ k-points meshes for relaxation and density of states calculations, respectively.[35] The [0001] surface of hexagonal IrSi$_3$ (with a space group of P6$_3$/mmc) was modeled by a slab with a thickness of 15 Å. The convergence criteria for the total energies were chosen as $10^{-5}$ eV between the consecutive ionic steps, and the maximum force allowed on each atom was set to 0.01 eV/ Å. Bottom layer Si atoms were terminated by hydrogen atoms in order to saturate the dangling bonds and we kept their positions fixed during structural relaxations. A vacuum region of 15 Å was introduced in order to prevent spurious interaction between the periodic slabs along z direction. Only the top three atomic layers were allowed to relax and reconstruct, and remaining layers were fixed at their corresponding bulk atomic positions in the calculations.

**Results & Discussion**

We first realized the formation of silicene like domains while studying the morphology of Ir modified Si(111) surface at relatively high Ir coverages (> 1ML). We choose to call our system as **S**ilicene-**L**ike **D**omain (SLD) and avoid calling it silicene without experimental proof showing that SLD's are electronically decoupled from the substrate. Figure 1a and 1b show large scale STM images of the surface with these islands. The islands were on different samples but grown under the same conditions. The islands are similar in size and shape however island in Figure 1b have SLD's whereas island in Figure 1a is flat and featureless. The wetting layer surrounding the island is made out of Ir-ring clusters that form $\sqrt{7} \times \sqrt{7}$ domains on Si(111). [36, 37] Ring clusters consists of a transition metal surrounded by six Si atoms, three of which directly bind to the transition metal and are called bridging adatoms, and the remaining three Si atoms, called the capping adatoms which bind two bridging adatoms and one Si substrate atom. Ir/Si(111) system exhibits Stranski–Krastanov type growth. The Ir-ring clusters function as a wetting layer, further deposition of Ir leads to the formation of the islands. STM image in Figure 1c shows a higher resolution picture of an SLD. The unit cell is indicated by a red rhombus enclosing total of six protrusions, three per half unit cell. The size of the unit cell is 23.1 Å.

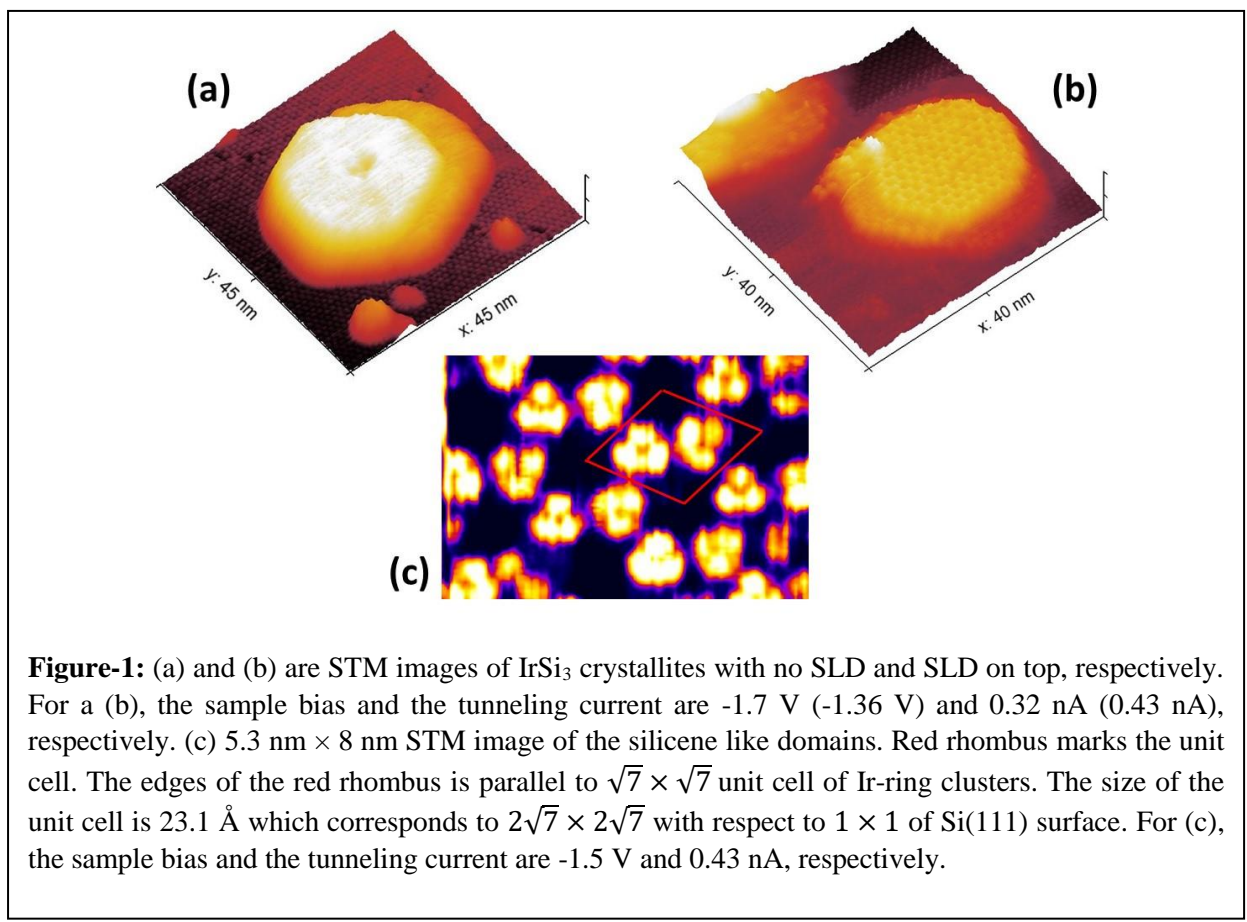

**Figure-1:** (a) and (b) are STM images of IrSi$_3$ crystallites with no SLD and SLD on top, respectively. For a (b), the sample bias and the tunneling current are -1.7 V (-1.36 V) and 0.32 nA (0.43 nA), respectively. (c) 5.3 nm × 8 nm STM image of the silicene like domains. Red rhombus marks the unit cell. The edges of the red rhombus is parallel to $\sqrt{7}\times\sqrt{7}$ unit cell of Ir-ring clusters. The size of the unit cell is 23.1 Å which corresponds to $2\sqrt{7}\times2\sqrt{7}$ with respect to $1\times1$ of Si(111) surface. For (c), the sample bias and the tunneling current are -1.5 V and 0.43 nA, respectively.

There are previous studies on bulk properties of iridium silicides epitaxially grown on various Si surfaces.[38, 39,40] Ir-silicides have three known crystal phases (IrSi, Ir$_3$Si$_5$ and IrSi$_3$), which can be grown selectively, and each phase grows in a specific temperature range. Among those phases, IrSi and IrSi$_3$ are metallic and Ir$_3$Si$_5$ is a semiconductor with a band gap of about 1.2 eV.[41] The phase diagram of bulk Ir-Si system shows eutectic at 1222 °C and at a composition of 80.5 % Si and 19.5 % Ir.[42] IrSi$_3$ phase is the phase closest to the bulk eutectic composition. When eutectic droplet cools down, it goes through spinodal decomposition and two stable phases emerge, Ir-free Si (forms the bulk Si) and IrSi$_3$. Due to relatively small Ir coverage on the surface (1 to 2 ML), the current system follows the phase diagram on the Si-rich side. However, the temperatures at which IrSi$_3$ forms are much lower than that of the bulk eutectic point. We attributed this to the fact that for nanostructures both solidus and liquidus curves shift to lower temperatures than that of the bulk phase diagram.[43,44] A similar shift of the eutectic point to lower temperatures has been observed in vacuum-liquid-solid growth of Ge nanowires when the catalyst was supersaturated.[45] Lin et. al. grew $IrSi_3$ on Si(111) at much lower temperature (ranging from 630 °C to 800 °C) than eutectic point by co-depositing Ir and Si on Si(111) in stoichiometric ratio.[46] They reported the growth of three major modes (A, B and C) of epitaxial silicides. Each mode of growth refer to a specific set of orientation relation between $IrSi_3$ and the Si(111) substrate.[47] Silicides grown at 630 °C exhibited two types (A and B) with approximately 50% of the area covered by each mode and 100% of the surface covered with Si. On the other hand, for $IrSi_3$ layers grown at 780 °C, epitaxial island formation of type C which cover approximately 70% of the surface was observed. Among these modes of growth, mode A and mode C are vicinal surfaces and cannot explain the formation of SLDs. However, mode B corresponds to the growth of [0001] of $IrSi_3$ parallel to [111] of Si. Based on Lin et. al.'s paper and the STM images we measured, we chose to use [0001] terminated surface of $IrSi_3$ for DFT calculations. (Figure 2) $IrSi_3$ has a hexagonal crystal structure with a=4.37 Å c= 7.40 Å.[48] In order to simulate Si-rich phase observed experimentally by Lin et.al., we started with Si terminated surface. Top of the surface comprises of two layers of Si with a height difference of 1.27 Å. (Figure 2a) In this model, we first relaxed the top three atomic layers of the [0001] surface of $IrSi_3$. Upon relaxation, these Si layers formed a silicene like structure with larger Si-Si bond lengths (around 2.4-2.6 Å) and smaller buckling (~0.1 Å) than that of a free standing silicene.[49] (Figure 2b) We also observed that Si atoms of silicene layer undergo a significant reconstruction in such a way that after relaxation the subsurface Ir atoms reside just below the midpoint of Si-Si bonds instead of being right under the Si atoms of the unreconstructed surface. (Figure 2c and 2d) This structural reconstruction helps Ir and Si atoms to saturate their dangling bonds and lower the surface energy.

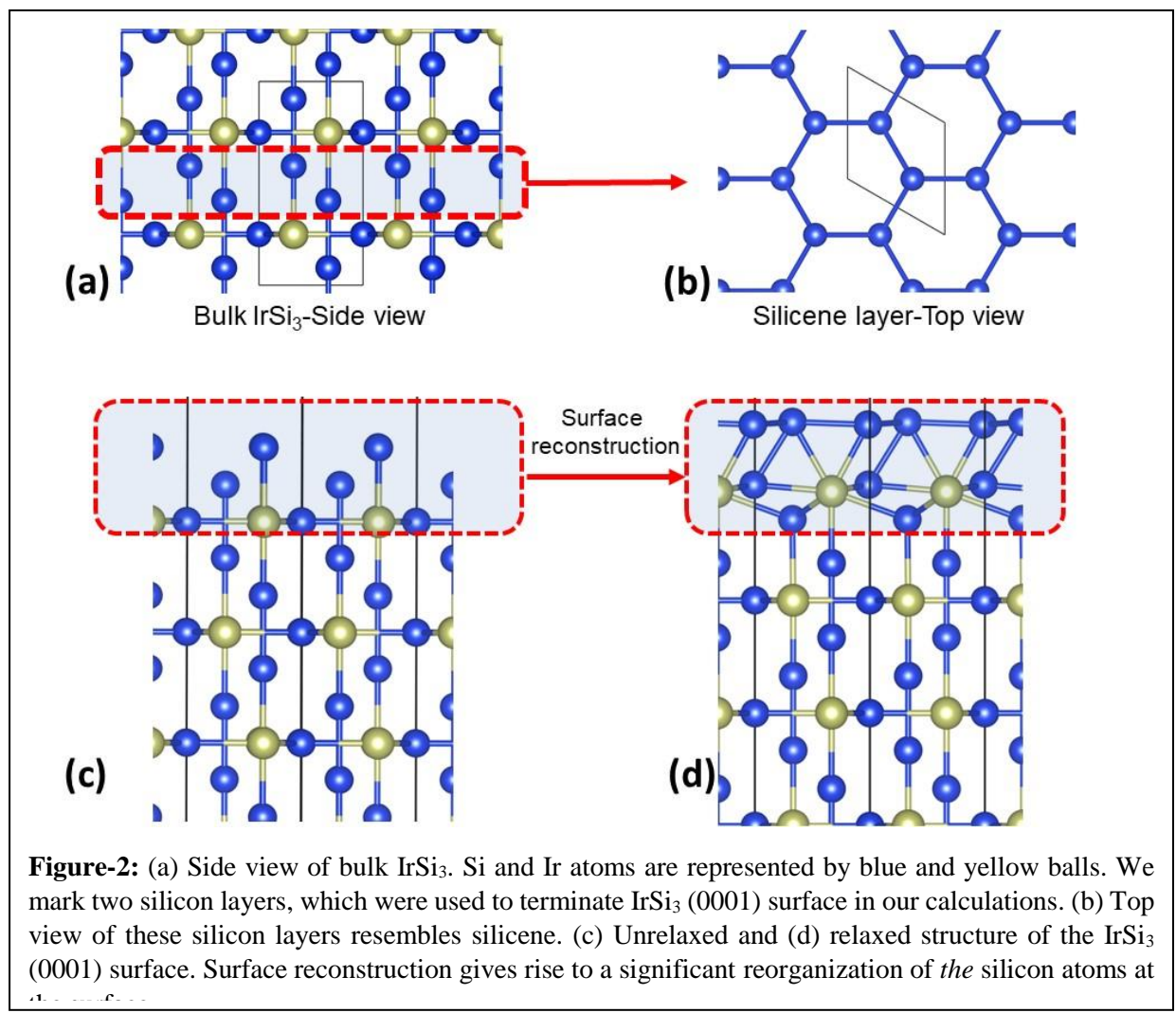

**Figure-2:** (a) Side view of bulk IrSi$_3$. Si and Ir atoms are represented by blue and yellow balls. We mark two silicon layers, which were used to terminate IrSi$_3$ (0001) surface in our calculations. (b) Top view of these silicon layers resembles silicene. (c) Unrelaxed and (d) relaxed structure of the IrSi$_3$ (0001) surface. Surface reconstruction gives rise to a significant reorganization of *the* silicon atoms at

High resolution STM images measured on SLDs show that there are six protrusions in the unit cell. The average distance between two nearest neighbor protrusions is about 0.6 nm which is similar to the distance between two Si adatoms on 7×7-Si(111) measured under similar tunneling conditions. Following the similarity with the adatoms of Si(111), we elevated some of the Si atoms by an amount of 0.5 Å so that the model of the surface reflects the symmetry observed in the STM images. However, when we relaxed the system, the elevated Si atoms returned to their original position. On the other hand, under biaxial strain, Si atoms stayed at the elevated positions suggesting that the observed surface reconstruction is a way to relieve built up strain. For the calculations presented in this paper, we fixed the elevated Si atoms and relaxed the unelevated Si atoms of SLDs. (Figure 3a and 3b) The lattice parameter of surface reconstruction of IrSi$_3$ (0001) is 23.1 Å. The edges of the red rhombus defining the unit cell is parallel to the $\sqrt{7} \times \sqrt{7}$ (with respect to $1 \times 1$ of Si(111)) unit cell of Ir-ring clusters formed on Si(111) surface. Therefore, the surface reconstruction of IrSi$_3$ (0001) can be described as $2\sqrt{7} \times 2\sqrt{7}$ (with respect to Si(111)). Upon relaxation, the height difference between the elevated and unelevated Si atoms becomes 0.48 Å, which is close to buckling of free standing silicene. Buckling creates a significant contrast in the simulated STM images. (Figure 3c)

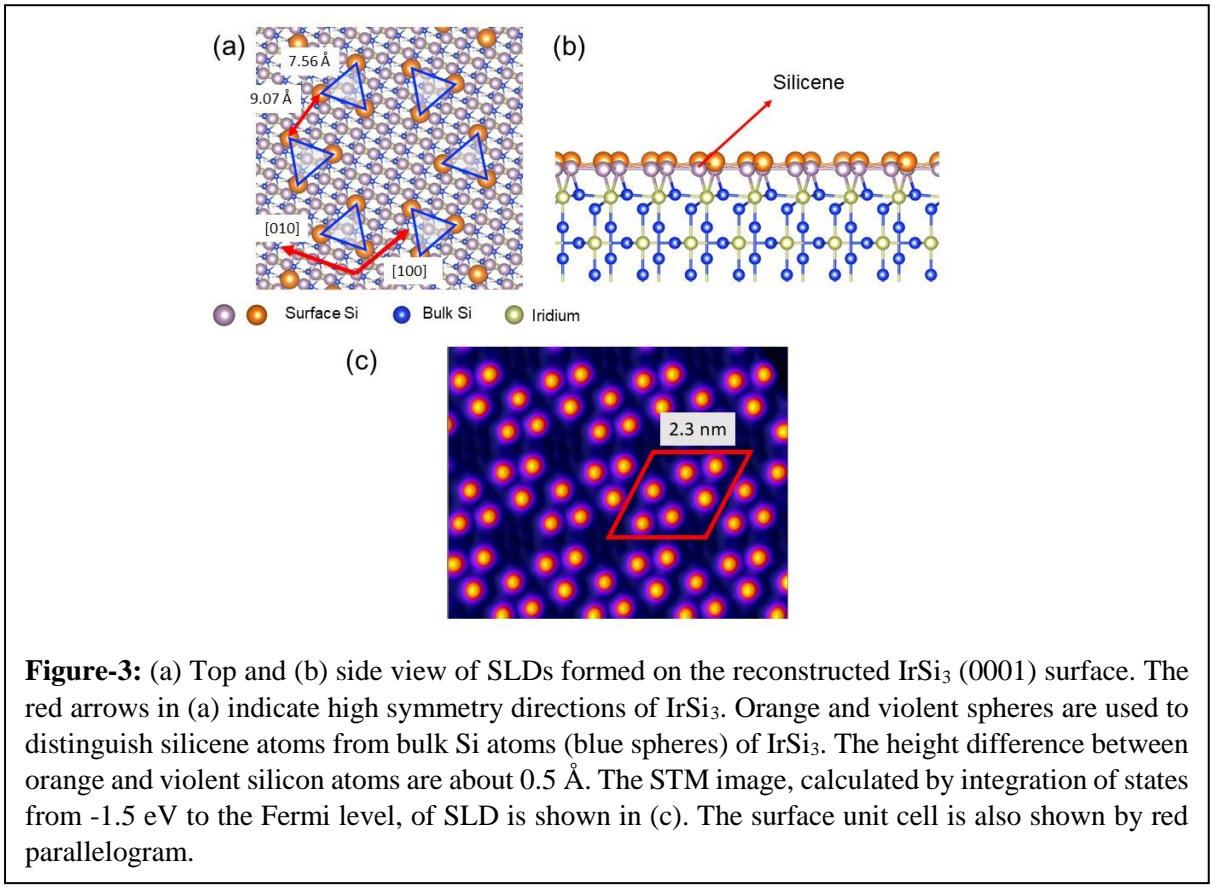

**Figure-3:** (a) Top and (b) side view of SLDs formed on the reconstructed IrSi$_3$ (0001) surface. The red arrows in (a) indicate high symmetry directions of IrSi$_3$. Orange and violent spheres are used to distinguish silicene atoms from bulk Si atoms (blue spheres) of IrSi$_3$. The height difference between orange and violent silicon atoms are about 0.5 Å. The STM image, calculated by integration of states from -1.5 eV to the Fermi level, of SLD is shown in (c). The surface unit cell is also shown by red parallelogram.

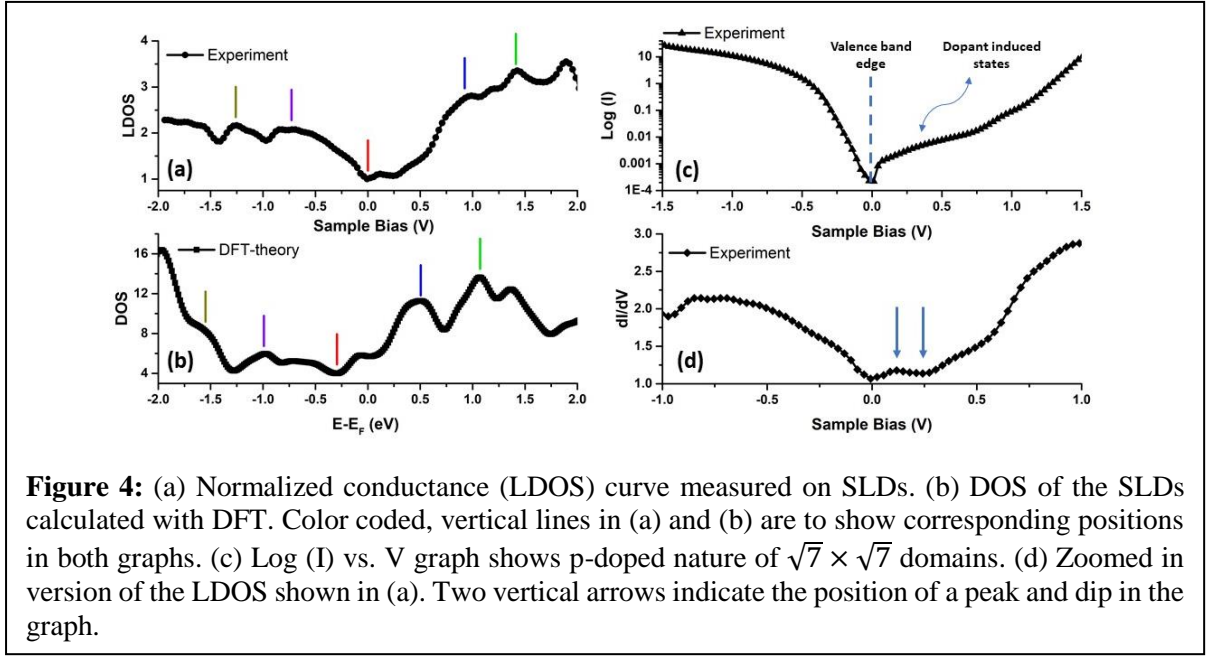

**Figure 4:** (a) Normalized conductance (LDOS) curve measured on SLDs. (b) DOS of the SLDs calculated with DFT. Color coded, vertical lines in (a) and (b) are to show corresponding positions in both graphs. (c) Log (I) vs. V graph shows p-doped nature of $\sqrt{7} \times \sqrt{7}$ domains. (d) Zoomed in version of the LDOS shown in (a). Two vertical arrows indicate the position of a peak and dip in the graph.

To reveal electronic properties of the SLDs, we performed STS measurements and calculated DOS with DFT. Both LDOS curves measured on SLDs and the calculated DOS curve exhibit similar features, however the dip measured around the Fermi level in LDOS curve is shifted to 0.3 eV below the Fermi level in the calculated DOS curve. (Figure 4a and 4b) A comparison between peak positions in both graphs also confirm a shift in similar magnitudes in the same direction. Due to the computational limitations, the DFT calculations were done only on $IrSi_3$ system, and the underlying Si bulk wasn't considered. $IrS_3$ islands are surrounded by $\sqrt{7} \times \sqrt{7}$ domains. (Figure 1a and 1b) The Log(I)-V curves measured on $\sqrt{7} \times \sqrt{7}$ domains show that the surface is p-type whereas the Si bulk is n-type.[50] (Figure 4c) This is only possible if the surface has acceptor type of states. Occupied acceptor states carry negative charges and induce upwards band bending.[51] If the band bending is strong enough due to high densities of surface acceptor states at lower energies in the forbidden band, there can be an inversion layer. For example, on Si(111)-7×7 surface, the surface states (acceptor type) cover almost all the bulk band gap (1.1 eV) and create an inversion layer. These surface states pin the Fermi level at about 0.7 eV above the valence band edge leading the formation of an electron depletion layer on n-type Si and a hole depletion layer for p-type Si. The band bending is about 0.4 eV. P-type nature of the $\sqrt{7} \times \sqrt{7}$ domains suggests occurrence of a situation in similar nature.

However, there is a difference between Si(111)-7×7 and $\sqrt{7} \times \sqrt{7}$ domains and that is the later has a band gap of about 0.75 eV indicating that the $\sqrt{7} \times \sqrt{7}$ domains have acceptor type of surface states localized closer to the bulk valence band edge. These acceptor states at the interface between $IrSi_3$ crystallites and the surface can acquire more electrons from IrSi3 crystallites to make the Fermi levels the same, leading to the observed downshift of the measured Fermi level of IrSi3 with respect to the one calculated.

The LDOS curve shown in Figure 4d is the same as the one in Figure 4a. We only plot between -1 V to 1 V to have a better visual around the Fermi level. Although the LDOS curve presented in Figure 4a and 4d don't exhibit any band gap as it was claimed to be present in the LDOS curves measured on silicene/$MoSi_2$ [Ref. 29], the LDOS curves of these two systems exhibit similar features. For example, there seems to be a small peak at 0.1 eV and there is a dip at 0.25 eV. The positions of both the peak and the dip are shifted to lower energies in our system. This can be due to the difference in the electronic properties of underlying crystallites.

**Conclusion:**

In this study, we demonstrated a novel method to grow SLDs on Ir modified Si(111) surface which can pave the way to integrate silicene into silicon based technologies. After depositing > 1 ML of Ir and annealing the surface at 750 °C, $IrSi_3$ crystallites form. Most of the crystallites have featureless morphology suggesting that they were terminated by Ir-rich surface of $IrSi_3$. We attribute this to the fact that as crystallites grow, Si atoms must come from the substrate through the crystallites to form the top layers which limits the supply of Si atoms. As a result of that most of the crystallites terminate with Ir-rich surface. dI/dV curves measured on these crystallites confirm that they are metallic. On the other hand, some $IrSi_3$ crystallites have SLDs grow on the top surface similar to a low-buckled silicene. The LDOS curves measured on the SLDs have non-zero density of states around the Fermi level indicating the metallic nature of the surface. There is a small dip at 0.25 eV. A similar feature observed on silicene/$MoSi_2$ and silicene/$ZrB_2$(0001) systems. The dip was suggested to be originating from a gap opening around the Dirac energy, however these claims, including ours, needs to be verified by other methods such as ARPES.

**Acknowledgement:** This work was partially supported by the National Science Foundation (DMR-1306101). Computer resources used in this work is provided by Computational Research Center (HPC Linux cluster) at University of North Dakota. A part of this work was supported by University of North Dakota Early Career Award (Grant number: 20622-4000-02624).